# Mapping a Hadamard Quantum Walk to a Unique Case of a Birth and Death Process

**Arie Bar-Haim**

Davidson Institute of Science Education, Weizmann Institute of Science, Rehovot 7610001, Israel; arik.bar-haim@weizmann.ac.il

**Abstract:** A new model maps a quantum random walk described by a Hadamard operator to a particular case of a birth and death process. The model is represented by a 2D Markov chain with a stochastic matrix, i.e., all the transition rates are positive, although the Hadamard operator contains negative entries (this is possible by increasing the dimensionality of the system). The probability distribution of the walker population is preserved using the Markovian property. By applying a proper transformation to the population distribution of the random walk, the probability distributions of the quantum states $|0>, 1>$ are revealed. Thus, the new model has two unique properties: it reveals the probability distribution of the quantum states as a unitary system and preserves the population distribution of the random walker as a Markovian system.

**1. Introduction**: Various problems of quantum random walks have been investigated by many groups in recent years. Among others, Aharonov et al. [1] explored quantum random walks, Ambainis et al. examined quantum walks on graphs [2–4], Bach et al. [5] investigated one-dimensional quantum walks with absorbing boundary conditions, Dür et al. [6] discussed quantum random walks in optical lattices, Konno et al. [7] examined absorption problems and the eigenvalues of two-state quantum walks [8], Mackay et al. [9] explored quantum walks in higher dimensions, and Bartlet et al. [10] examined quantum topology identification in addition to various other problems [11,12]. Several other studies discussed the differences between random walks and quantum random walks, such as those by Childs et al. [13] and Motes et al. [14].

In the present study, a new model that maps a quantum random walk described by a Hadamard operator to a particular case of a random walk is presented. The model is represented by a Markov chain with a stochastic matrix, i.e., all the transition rates are positive, although the Hadamard operator contains negative entries. This is possible by increasing the dimensionality of the system. The probability distribution of the walker population is preserved by the Markovian property, and by applying a proper transformation to the population distribution of the random walk, the probability distributions of the quantum states $|0>, 1>$ are revealed as if they were a unitary system. Thus, the model presented here has two unique properties: a probability distribution of the quantum states as a unitary system and a population distribution of the random walker as a Markovian system.

**2. Mapping the Hadamard Operator to a 4 × 4 Symmetric Markov Chain**

The discrete-time quantum random walk is defined by two operators [2,3], namely the Hadamard operator that flips the quantum state of each site, and the shift operator, which moves the quantum states depending on whether the quantum states are $|0>$ or $|1>$.

The Hadamard matrix, $H$, is defined by the following unitary operator:

$$H = \frac{1}{\sqrt{2}}\begin{bmatrix} 1 & 1 \\ 1 & -1 \end{bmatrix}, \tag{1}$$

and the coin states can be defined by $|0>$ and $|1>$ as follows:

$$|0>= \begin{pmatrix} 1 \\ 0 \end{pmatrix} \quad |1>= \begin{pmatrix} 0 \\ 1 \end{pmatrix}, \qquad (2)$$

The Hadamard operator can be mapped by four states of a Markov chain, as shown in Figure 1 [15], with transition probabilities of $p = q = 0.5$.

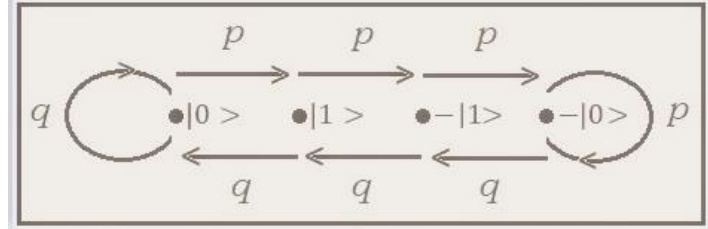

**Figure 1.** A Markov chain representing the transitions between four states. Note that each site is named by the following symbols: $|0>$, $|1>$, $-|1>$, $-|0>$.

This can be shown explicitly by the following matrix product:

$$H = \frac{1}{\sqrt{2}} BAB^T \qquad (3)$$

where $A$ is the transition probability matrix of the system described in Figure 1:

$$A = \begin{bmatrix} 0.5 & 0.5 & 0 & 0 \\ 0.5 & 0 & 0.5 & 0 \\ 0 & 0.5 & 0 & 0.5 \\ 0 & 0 & 0.5 & 0.5 \end{bmatrix}, \qquad (4)$$

and matrix $B$ is defined as

$$B = \begin{bmatrix} 1 & 0 & 0 & -1 \\ 0 & 1 & -1 & 0 \end{bmatrix}, \qquad (5)$$

note that $\text{Det}(H) \neq 0$, while $\text{Det}(A) = 0$.

This mapping will be used in the following sections.

### 3. Building an RW Model Using the Four-Site Chain in the $y$-axis

The second operator of the Hadamard walk, is the shift operator [2], which propagates the quantum states to the right or left, depending on whether the quantum states are $|0>$ or $|1>$. Thus, it actually is a birth and death random walk process. Combining the shift operator with the Markov chain described by the matrix A yields the RW model depicted in Figure 2. The horizontal direction, the $x$-axis, describes the movements due to the shift operator, and the vertical direction, the $y$-axis, represents the movements due to the Hadamard operator.

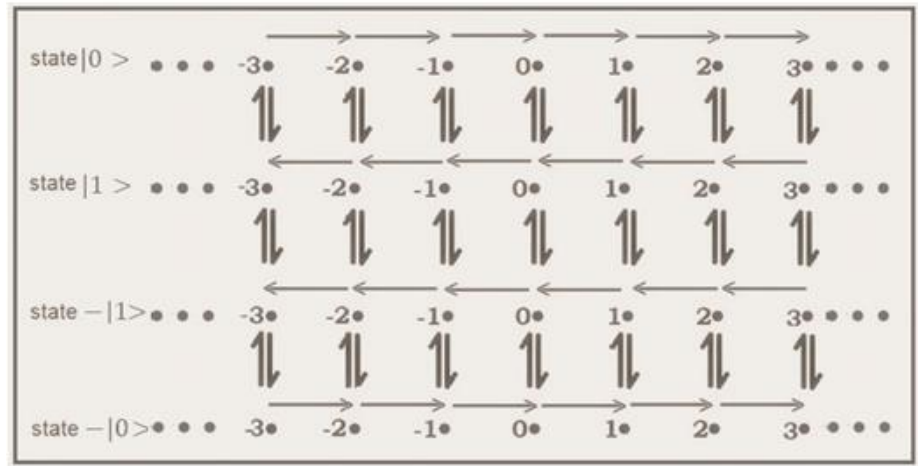

**Figure 2.** A two-step RW model. Note that the first to fourth rows are named as: $|0>$, $|1>$, $-|1>$, $-|0>$.

The main points of the RW model are as follows:

1. The shift operator is responsible for the movement in the horizontal direction ($x$-axis).
2. The transition matrix $A$ is responsible for the movement in the vertical direction ($y$-axis).
3. The movements in each direction occur one after the other, rather than simultaneously.

Similarly, the Hadamard walk can also be presented as described in Figure 3:

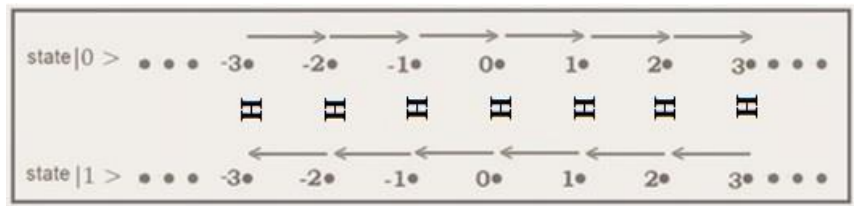

**Figure 3.** A two-step QRW model.

The main points of the QRW model are as follows

1. The shift operator is responsible for the movement in the horizontal direction ($x$-axis).
2. The transition matrix $H$ is responsible for the movement in the vertical direction ($y$-axis).
3. The movements in each direction occur one after the other, rather than simultaneously.

**Table 1.** presents a mathematical comparison between these models (for an infinite case).

|  | **Hadamard Walk** | **Random Walk Model** |
|---|---|---|
| $Y$-axis operator—Hadamard operator in space | $y = I_d \otimes H$ | $Y = I_d \otimes A$ |
| $X$-axis operator—Shift operator in space | $x = Right \otimes Zero + Left \otimes One$ | $X = Right \otimes \hat{Z}ero + Left \otimes \hat{O}ne$ |
| The dynamics of the whole process | $u = xy =$ $Right \otimes ZeroH + Left \otimes OneH$ | $U = XY =$ $Right \otimes \hat{Z}eroA + Left \otimes \hat{O}neA$ |

The table describes the Hadamard walk dynamics versus the RW model. The equivalence dynamics between these models are proved in detail in Appendixes A and B and in [15]:

$$u(I_d \otimes B) = (I_d \otimes B)(\sqrt{2}U) \tag{6}$$

And after $n$ steps

$$u^n(I_d \otimes B) = (\sqrt{2})^n (I_d \otimes B) U^n \tag{7}$$

Note that $u^n$ and $U^n$ represent the dynamics after n steps of the Hadamard walk and of the QRW, correspondingly.

Where: $I_d$ is a unit matrix of size $d$

$$Zero = \begin{bmatrix} 1 & 0 \\ 0 & 0 \end{bmatrix}, \; One = \begin{bmatrix} 0 & 0 \\ 0 & 1 \end{bmatrix} \tag{8}$$

$$\hat{Z}ero = \begin{bmatrix} 1 & 0 & 0 & 0 \\ 0 & 0 & 0 & 0 \\ 0 & 0 & 0 & 0 \\ 0 & 0 & 0 & 1 \end{bmatrix}, \; \hat{O}ne = \begin{bmatrix} 0 & 0 & 0 & 0 \\ 0 & 1 & 0 & 0 \\ 0 & 0 & 1 & 0 \\ 0 & 0 & 0 & 0 \end{bmatrix} \tag{9}$$

The *Right* and the *Left* matrices are $d \times d$ zero matrices, except those entries that appear above and beneath the main diagonal. Specifically:

$Right\,(j, j+1) = 1$ and $Left(j+1, j) = 1$ for any integer $j \in (1, d-1)$ as follows:

$$Right = \begin{bmatrix} 0 & 1 & 0 & 0 & 0 & \ldots \\ 0 & 0 & 1 & 0 & 0 & \ldots \\ 0 & 0 & 0 & 1 & 0 & \ldots \\ 0 & 0 & 0 & 0 & 1 & \ldots \\ 0 & 0 & 0 & 0 & 0 & \ldots \\ \ldots & \ldots & \ldots & \ldots & \ldots & \ldots \end{bmatrix}, Left = \begin{bmatrix} 0 & 0 & 0 & 0 & 0 & \ldots \\ 1 & 0 & 0 & 0 & 0 & \ldots \\ 0 & 1 & 0 & 0 & 0 & \ldots \\ 0 & 0 & 1 & 0 & 0 & \ldots \\ 0 & 0 & 0 & 1 & 0 & \ldots \\ \ldots & \ldots & \ldots & \ldots & \ldots & \ldots \end{bmatrix} \tag{10}$$

The last manipulation that is involved is multiplying Equation (7) by the initial condition.

The initial condition of the system described by Figure 2. has the following mathematical presentation for $n = 0$:

$$P(n) = \begin{bmatrix} p_{|0>,1} & p_{|0>,2} & p_{|0>,3} & p_{|0>,4} & \cdots & p_{|0>,d} \\ p_{|1>,1} & p_{|1>,2} & p_{|1>,3} & p_{|1>,4} & \cdots & p_{|1>,d} \\ p_{-|1>,1} & p_{-|1>,2} & p_{-|1>,3} & p_{-|1>,4} & \cdots & p_{-|1>,d} \\ p_{-|0>,1} & p_{-|0>,2} & p_{-|0>,3} & p_{-|0>,4} & \cdots & p_{-|0>,d} \end{bmatrix} \tag{11}$$

Note that the first subscript describes the row and the second subscript represents the site number. Since a Kronecker product is involved, the initial condition should be changed into a column vector (vectorization) as follows:

$$M = \begin{bmatrix} a & c & e \\ b & d & f \end{bmatrix} \xrightarrow{yields} M = [a, b, c, d, e, f]^T$$

The vectorization of $P(0)$ is as follows:

$$\widetilde{P(0)} = [p_{|0>,1}, p_{|1>,1}, p_{-|1>,1}, p_{-|0>,1}, p_{|0>,2}, p_{|1>,2}, p_{-|1>,2}, p_{-|0>,2} \cdots \cdots p_{-|0>,d}]^T, \tag{12}$$

Using the initial condition and taking the square of the absolute yields:

$$V = \left| u^n (I_d \otimes B) \widetilde{P(0)} \right|^2 = 2^n \left| (I_d \otimes B) U^n \widetilde{P(0)} \right|^2 \tag{13}$$

$V$ is a column vector whose odd entries present the probability distribution of the |0> state and its even entries give the probability distribution of the |1>.

Thus:

$$\begin{bmatrix} |\psi_{|0>}(n)|^2 \\ |\psi_{|1>}(n)|^2 \end{bmatrix} = \begin{bmatrix} V(a) \\ V(b) \end{bmatrix} \tag{14}$$

where $a = 1, 3, 5, \ldots 2d - 1; b = 2, 4, 6, \ldots 2d.$ and $|\psi_{|0>}(n)|^2, |\psi_{|1>}(n)|^2$ describes the probability distribution in space of the quantum states $|0>$ and $|1>$, respectively.

By defining $\tilde{V} = U^n \widehat{P(0)}$, the matrix $P(n)$ described in Equation (11) can be re-generated as four-row vectors as follows:

$$P_{|0>}(n) = \tilde{V}(a), P_{|1>}(n) = \tilde{V}(b), P_{-|1>}(n) = \tilde{V}(c), P_{-|0>}(n) = \tilde{V}(d) \tag{15}$$

where $a = 1, 5, 9, \ldots 2d - 1, b = 2, 6, 10 \ldots 4d - 2, c = 3, 7, 11 \ldots 4d - 3, d = 4, 8, 12 \ldots 4d$ and

$$P(n) = \begin{bmatrix} P_{|0>}(n) \\ P_{|1>}(n) \\ P_{-|1>}(n) \\ P_{-|0>}(n) \end{bmatrix} \tag{16}$$

Thus, the probability distribution of quantum states $|0>$ and $|1>$ after $n$ steps in space can also be written, using Equations (14) and (15), in a simple way as:

$$|\psi_{|0>}(n)|^2 = 2^n |P_{|0>}(n) - P_{-|0>}(n)|^2$$
$$|\psi_{|1>}(n)|^2 = 2^n |P_{|1>}(n) - P_{-|1>}(n)|^2 \tag{17}$$

Note that $P_{|0>}(n)$ is the first row of the matrix, $P(n)$, $P_{|1>}(n)$ is the second row, $P_{-|1>}(n)$ is the third row, and $P_{-|1>}(n)$ is the last row. In other words, a simple manipulation of the RW distribuition yields the probability distribution of the |0> and |1> quantum states.

Figure 4. describes the same results for the two models after 20 steps, beginning at the 40th site and initializing with $i|0>$ for a system with 80 sites (The complete Matlab program is presented in Appendix D).

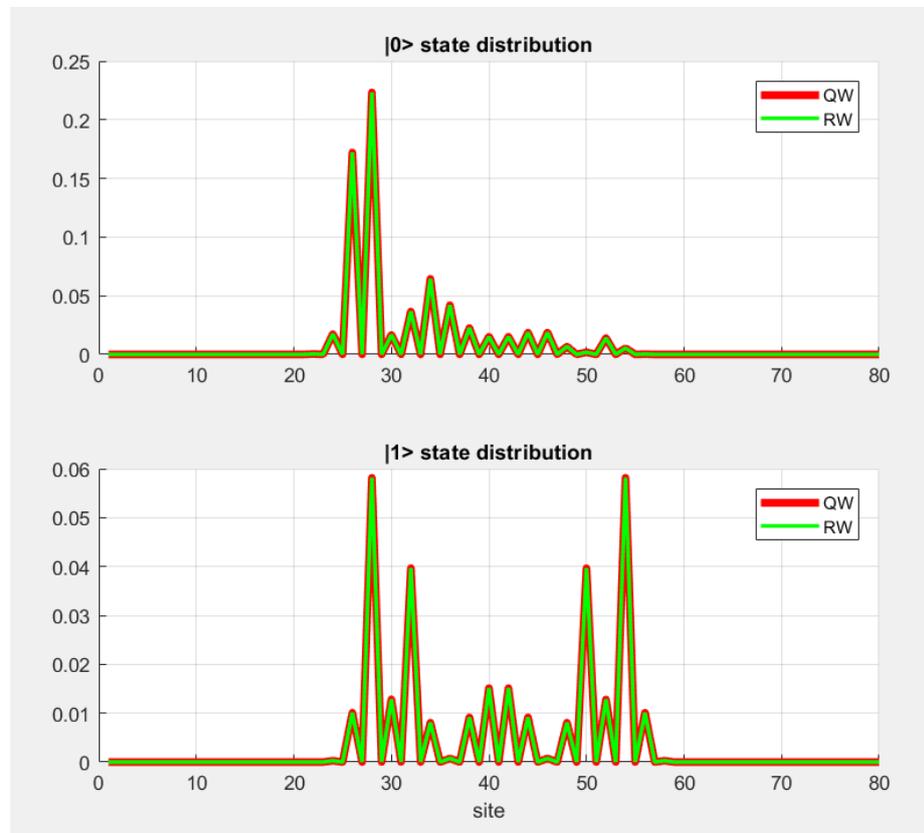

**Figure 4.** Comparison between RW and Hadamard walk after 20 steps beginning at the 40th site and initializing with $i|0>$ for a system with 80 sites.

The next six graphs, Figure 5, present in detail the RW system, starting at the origin with $|0>$ after 100 steps. Each row of the matrix describing the RW process yields Gaussian population distributions, and when interfering with these distributions (Equation (17)), the probability distribution of $|0>$ and $|1>$ are revealed.

The following graphs are:

(a) RW population distribution of the positive $|0>$ state in space (the first row of P(n)) after 100 steps, represented by $P_{|0>}(n)$ (Note the Gaussian distribution).
(b) RW population distribution of the negative $|0>$ state in space (The last row of P(n)) after 100 steps, described by $P_{-|0>}(n)$ (Note the Gaussian distribution).
(c) Energy distribution of the $|0>$ quantum state, which is obtained by interfering with these rows and taking the absolute value multiplied by $2^n$, corresponding to Equation (13)

The next three graphs are similar:

(d) RW population distribution of the positive $|1>$ state (The second row of P(n)), described by $P_{|1>}(n)$.
(e) RW population distribution of the negative $|1>$ state (The third row of P(n)), described by $P_{-|1>}(n)$.
(f) Energy distribution of the $|1>$ quantum state.

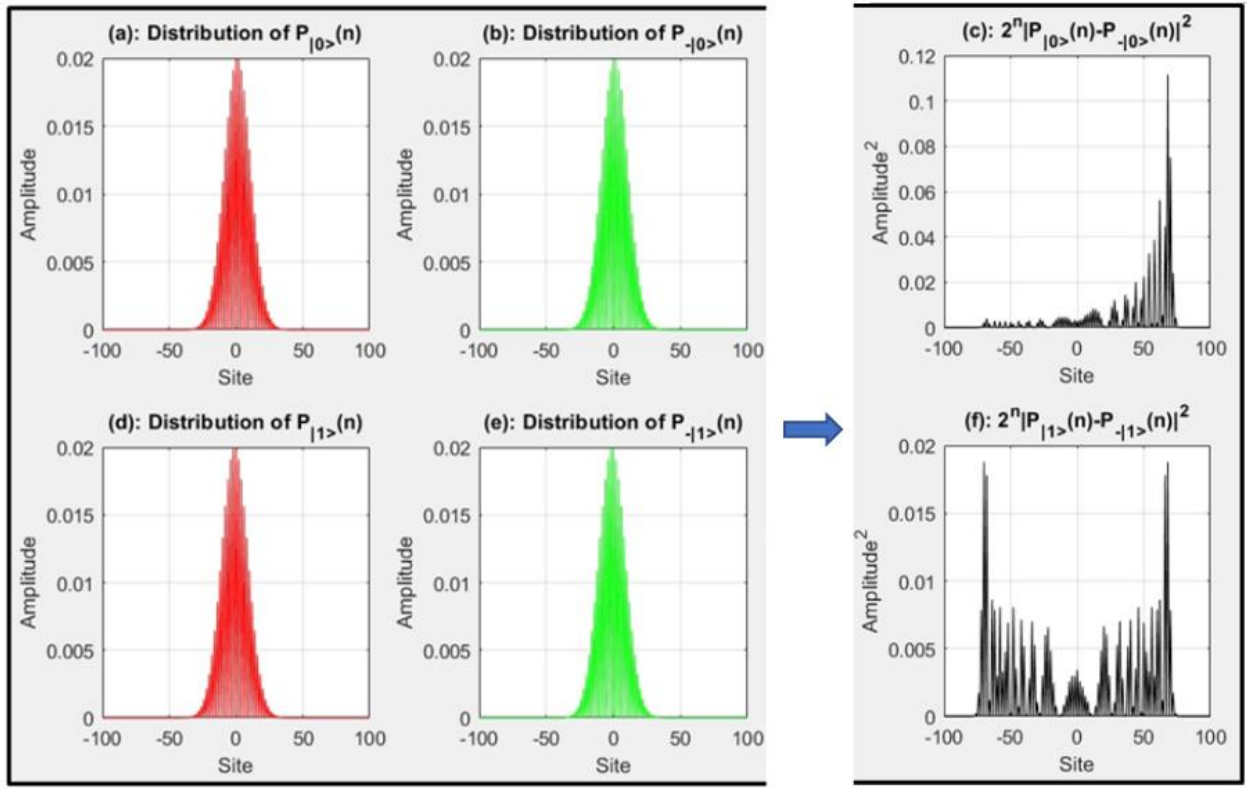

**Figure 5.** Graphs (**a**,**b**) present the RW population distribution of $|0>$ (the first row of P(n)) and $-|0>$ (The last row of P(n)). On this basis, chart (**c**) presents the probability distribution of the quantum state $|0>$. The graphs in the second row are the same in respect of state $|1>$.

## 4. Summary

A new model that maps a Hadamard walk to a birth and death case of a RW model was presented here. The RW walk model has completely different properties. For example, an infinite system has a stochastic matrix; namely, the system conserves the random walk population. Using a transformation applied to the population distribution of the RW model, the quantum state distribution of the Hadamard system was revealed. Thus, the model has two unique properties: conservation of the quantum distribution as a unitary system and of the population distribution as a Markovian system.

**Data Availability Statement:** https://arxiv.org/abs/2104.04286

**Conflicts of Interest:** "The authors declare no conflict of interest."

**Appendix A**

The dynamics of the Hadmarad walk can be written as $= xy$, where according to Table 1 Equaiton.

$$y = I_d \otimes H \tag{A1}$$

$$x = Right \otimes Zero + Left \otimes One \tag{A2}$$

$$u = (Right \otimes Zero + Left \otimes One)\, I_d \otimes H = Right \otimes ZeroH + Left \otimes OneH \tag{A3}$$

The dynamic of the Markov chain depicted in Figure 2 is formulated similarly:

$$Y = I_d \otimes A \tag{A4}$$

and the $x$-axis operators can be written as:

$$X = Right \otimes \hat{Z}ero + Left \otimes \hat{O}ne \tag{A5}$$

The dynamic of a complete step of the RW system is:

$$U = XY = Right \otimes \hat{Z}eroA + Left \otimes \hat{O}neA \tag{A6}$$

The equivalence between the two dynamics is proven step-by-step, beginning by substituting
$H = \frac{1}{\sqrt{2}} BAB^T$ into Equation (A1), which obtains:

$$y = I_d \otimes H = \frac{1}{\sqrt{2}}(I_d \otimes BAB^T) \tag{A7}$$

The shift operator $x$, described by Equation (A2), can be written in terms of the matrices $\hat{Z}ero, \hat{O}ne, A, B$, using the following relationship:
$Zero = \frac{1}{2} B\hat{Z}eroB^T$ and $One = \frac{1}{2} B\hat{O}neB^T$
which can be shown explicitly by multiplication. Substituting these relationships into Equation (A2) yields:

$$x = Right \otimes Zero + Left \otimes One = \frac{1}{2}\left(Right \otimes B\hat{Z}eroB^T + Left \otimes B\hat{O}neB^T\right) \tag{A8}$$

Therefore:

$$u = xy = \frac{1}{2}(Right \otimes B\hat{Z}eroB^T + Left \otimes B\hat{O}neB^T)\,\frac{1}{\sqrt{2}}(I_d \otimes BAB^T) \tag{A9}$$

Using the Kronecker product property of $(a \otimes b)(c \otimes d) = ac \otimes bd$ on Equation (A9) yields

$$u = \frac{1}{2\sqrt{2}}(Right \otimes B\hat{Z}eroB^T BAB^T + Left \otimes B\hat{O}neB^T B\, AB^T) \tag{A10}$$

Multiplication of both sides of Equation (A10) by $(I_d \otimes B)$ yields:

$$u(I_d \otimes B) = \frac{1}{2\sqrt{2}}(Right \otimes B\hat{Z}eroB^T BAB^T B + Left \otimes B\hat{O}neB^T B\, AB^T B) \tag{A11}$$

since $[\hat{Z}ero, B^T B] = 0$, $[\hat{O}ne, B^T B] = 0$ and $[A, B^T B] = 0$ (simply by multiplication)

Consequently, Equation (A11) can be rearranged as follows:

$$u(I_d \otimes B) = \frac{1}{2\sqrt{2}}(Right \otimes BB^T BB^T B\hat{Z}eroA + Left \otimes BB^T BB^T B\hat{O}neA) \quad (A12)$$

Now, using the properties of $BB^T = 2I_2$ yields:

$$u(I_d \otimes B) = \sqrt{2}(Right \otimes B\hat{Z}eroA + Left \otimes B\hat{O}ne\,A) \quad (A13)$$

and rearranging the last equation using the Kronecker product property yields:

$$u(I_d \otimes B) = \sqrt{2}(I_d \otimes B)(Right \otimes \hat{Z}eroA + Left \otimes \hat{O}neA) \quad (A14)$$

Thus,

$$u(I_d \otimes B) = \sqrt{2}(I_d \otimes B)U \quad (A15)$$

**Appendix B**

Multiplying both sides of Equation (A15) by $I_d \otimes B^T$ yields:

$$u = \frac{1}{\sqrt{2}}(I_d \otimes B)U(I_d \otimes B^T) \quad (A16)$$

As a consequence, $u^2$ can be formulated explicitly as:

$$u^2 = \frac{1}{\sqrt{2}}(I_d \otimes B)U(I_n \otimes B^T)\frac{1}{\sqrt{2}}(I_d \otimes B)U(I_d \otimes B^T) \quad (A17)$$

Next, multiplying both sides by $I \otimes B$ yields:

$$u^2(I_d \otimes B) = \frac{1}{\sqrt{2}}(I_d \otimes B)U(I_d \otimes B^T)\frac{1}{\sqrt{2}}(I_d \otimes B)U(I_d \otimes B^T)(I_d \otimes B) \quad (A18)$$

and since $U$ commutes as $U(I_d \otimes B^T)(I_d \otimes B) = (I_d \otimes B^T)(I_d \otimes B)U$ (see Appendix C) and $(I_d \otimes B)(I_d \otimes B^T) = I_d \otimes BB^T = 2(I_d \otimes I_2)$, then:

$$u^2(I_d \otimes B) = \left(\sqrt{2}\right)^2 (I_d \otimes B)U^2 \quad (A19)$$

And in general

$$u^n(I_d \otimes B) = \left(\sqrt{2}\right)^n (I_d \otimes B)U^n \quad (A20)$$

**Appendix C**

$$U = Right \otimes \hat{Z}eroA + Left \otimes \hat{O}neA \quad (A21)$$

By using the Kronecker product property:

$$U(I_d \otimes B^T)(I_d \otimes B) = Right \otimes \hat{Z}eroAB^T B + Left \otimes \hat{O}neAB^T B \quad (A22)$$

$$(I_d \otimes B^T)(I_d \otimes B)U = Right \otimes B^T B\hat{Z}eroA + Left \otimes B^T B\hat{O}neA \quad (A23)$$

Since $[B^T B, \hat{Z}ero] = 0 \text{ and } [B^T B, A] = 0$ explicitly by multiplication thus:

$$U(I_d \otimes B^T)(I_d \otimes B) = Right \otimes B^T B\hat{Z}eroA + Left \otimes B^T B\hat{O}neA \quad (A24)$$

Therefore: $U(I_d \otimes B^T)(I_d \otimes B) = (I_d \otimes B^T)(I_d \otimes B)U \quad (A25)$

## Appendix D. Matlab Program

```
% Comparison between a Hadamard Walk and the RW Model %
presented in the paper.
clear; close all
d=80; % The number of sites.
n=20;% The number of steps.
center=floor(d/2);% Center site.
P0(4,d)=0; % the initial condition of the RW.
P0(1,center)=1+1i/2 ;
P0(4,center)=1-1i/2 ;
%pwave_RW=vec(P0), vectorization of P0 generates %column vector
pwave_RW(4*d,1)=0;
pwave_RW(1:4:end)=P0(1,:);
pwave_RW(2:4:end)=P0(2,:);
pwave_RW(3:4:end)=P0(3,:);
pwave_RW(4:4:end)=P0(4,:);
% The initial condition of the Hadamard system.
p0(2,d)=0;
p0(1,:)=P0(1,:)-P0(4,:);
p0(2,:)=P0(2,:)-P0(3,:);
%Vectorization of p0 generates column vector %pwave_QW
```

```matlab
pwave_QW(2*d,1)=0;
pwave_QW(1:2:end)=p0(1,:);
pwave_QW(2:2:end)=p0(2,:);

%%% Hadamard walk system %%%
H=1/sqrt(2)*[1 1; 1 -1];%The Hadamard matrix
zero=[1 0 ; 0 0];
one=[0 0 ; 0 1];
Right(d,d)=0;Right(1:d-1,2:d)=eye(d-1);Left=Right';
x=kron(Right,zero)+kron(Left,one);y=kron(eye(d),H);u=x*y;
p=u^n*pwave_QW;
pzero=(abs(p(1:2:end))).^2;%|0> quantum state
pone= (abs(p(2:2:end))).^2;%|1> quantum state
%%% RW walk system %%%
p=0.5;
%Transition matrix A
A=[p p 0 0;p 0 p 0;0 p 0 p;0 0 p p];
Zero=[1 0 0 0;0 0 0 0; 0 0 0 0; 0 0 0 1];
One= [0 0 0 0; 0 1 0 0; 0 0 1 0; 0 0 0 0];
X=kron(Right,Zero)+kron(Left,One);Y=kron(eye(d),A);U=X*Y;
P=U^n*pwave_RW;% P(N)
```

```matlab
Pzero=2^n*(abs(P(1:4:end)-P(4:4:end))).^2;
Pone= 2^n*(abs(P(2:4:end)-P(3:4:end))).^2;
%%% Plotting %%%
subplot(2,1,1);hold on
plot(1:d,pzero,'r','linewidth',4);%QW model.
plot(1:d,Pzero,'g','linewidth',2);%RW lodel.
title ('|0> state distribution');grid on;legend('QW','RW');
subplot(2,1,2);hold on
plot(1:d,pone,'r','linewidth',4);%QW model.
plot(1:d,Pone,'g','linewidth',2);%RW model.
xlabel('site');
title ('|1> state distribution');grid on;legend('QW','RW');
%%%Display
energy=sum(Pzero+Pone)%Energy conservation
pop=sum(sum(P)) %Population conservation
strcat('energy=',num2str(energy),',population=',num2str(pop))
```